# The QuarkNet/Grid Collaborative Learning e-Lab

Marjorie Bardeen
*Fermilab*
mbardeen@fnal.gov

Eric Gilbert
*Fermilab*
egilbert@fnal.gov

Thomas Jordan
*Fermilab*
jordant@fnal.gov

Paul Nepywoda
*Fermilab*
nepywoda@fnal.gov

Elizabeth Quigg
*Fermilab*
liz@fnal.gov

Mike Wilde
*Argonne National Lab*
wilde@mcs.anl.gov

Yong Zhao
*University of Chicago*
yongzh@cs.uchicago.edu

## Abstract

*We describe a case study that uses grid computing techniques to support the collaborative learning of high school students investigating cosmic rays. Students gather and upload science data to our e-Lab portal. They explore those data using techniques from the GriPhyN collaboration. These techniques include virtual data transformations, workflows, metadata cataloging and indexing, data product provenance and persistence, as well as job planners. Students use web browsers and a custom interface that extends the GriPhyN Chiron portal to perform all of these tasks. They share results in the form of online posters and ask each other questions in this asynchronous environment. Students can discover and extend the research of other students, modeling the processes of modern large-scale scientific collaborations. Also, the e-Lab portal provides tools for teachers to guide student work throughout an investigation.*

*http://quarknet.uchicago.edu/elab/cosmic*

## 1. Introduction

Cosmic rays have been the subject of scientific research since 1912 when Victor Hess confirmed their existence from a balloon several kilometers above Earth. Our upper atmosphere is awash with these particles that are created in many astronomical objects. Most have low energies but still penetrate to Earth's surface. More fascinating are cosmic rays with energies six times higher than current theory allows. Two such events have prompted scientists to study more of these high-energy particles and to try to discover their source. [1] High school students have an opportunity to contribute to this research.

We are deploying a network of classroom cosmic ray detectors in high schools across North America. [2] The detectors capture the time and location of the arrival of cosmic rays and save the data to a local computer (Figure 1). They use GPS timing to ensure precise timestamps. [3] These detectors are sensitive enough to capture cosmic rays that fill in the energy spectrum from low to high and possibly even to see one of the rare high-energy events.

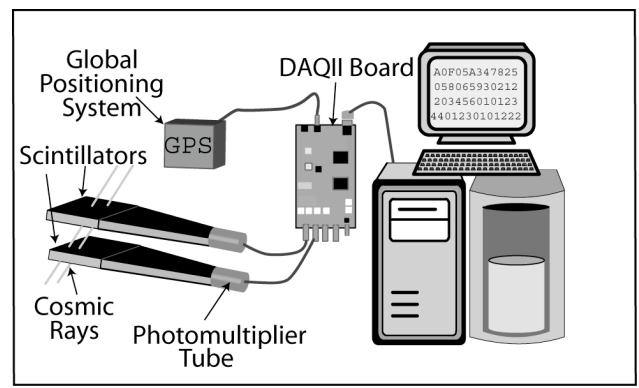

**Figure 1. The school-based detector**

In addition to conducting individual studies for muon flux or lifetime, students are able to set up detectors to function as an array that covers an area including several schools. When the local data are uploaded to a central server, students in the collaboration can investigate data from one school or combine data from multiple sites. Students without detectors can also be active members of the collaboration by accessing and analyzing data.

The science lends itself to project-based learning, a system for organizing portions of the curriculum around ill-structured problems that help students acquire knowledge while building problem-solving skills. The science calls for collaboration, requiring students to function in an experimental environment similar to high-energy physics collaborations, in which the web plays a vital part in supporting research. They must participate in a network of users that needs access to computer cycles, storage capacity, shared data and data products, community-defined analysis tools, and communication tools to share and discuss results. The QuarkNet/Grid Collaborative Learning e-Lab provides this online, cross-platform and ubiquitous environment.

## 2. Case study

For this case study we make an assumption that some research experiments using grids will have data interesting to and appropriate for secondary school classroom investigations, and that partnerships of computer scientists, research scientists and educators can create online collaborative learning environments, **e-Labs**, for students. The foundation of an e-Lab is the grid infrastructure and virtual data tools and techniques; experiments provide data, analysis software and background material; on top of these resources educators create user-friendly, single-purpose, application-specific web pages.

Our case study is a partnership between QuarkNet [4] and the GriPhyN [5] Project. QuarkNet is an education program for US ATLAS and USCMS, United States collaborations working on experiments for the Large Hadron Collider at the European Organization for Nuclear Research. Participants also work on as many as nine other high-energy experiments, for example, BaBar, CDF, DØ, FAST and MINOS. A long-term program, QuarkNet brings high school teachers and their students into research groups nationwide. Students learn fundamental physics as they analyze live online data and participate in inquiry-oriented investigations.

The GriPhyN Project is developing grid technologies for scientific and engineering projects that must collect and analyze distributed, petabyte-scale datasets. GriPhyN research will enable the development of Petascale Virtual Data Grids (PVDGs) through its Virtual Data Toolkit. [6]

The QuarkNet/Grid research project explores the potential of using virtual data grid tools and techniques to support student data-analysis projects for US ATLAS and USCMS. We begin with a pilot cosmic ray study that provides a chance for:
- Students to do authentic research—to access, process and publish data, report and share their results as online posters, and have online discussions with one another about their work.
- Student researchers to experience the environment of scientific collaborations.
- Student researchers to make contributions to a burgeoning scientific field.
- Educational researchers to evaluate the effectiveness of such an endeavor.

## 3. Collaborative learning environment

In our vision for collaborative learning, student research teams conduct investigations of real-world problems or issues through online projects that are collaborative, student-driven and technology-dependent. When skillfully applied, technology can enhance learning in new and powerful ways such as allowing students to reach beyond classroom walls to collaborate and publish original work to a worldwide audience. When skillfully designed, projects can require collaboration among student groups and between students and experts. With appropriate management tools, students can design and carry out an investigation even though they are novice researchers, and teachers can facilitate student learning, track their progress and assess their work.

We share a vision for technology-enhanced learning with such projects as EleGI and GRIDCOLE. Because we want our students to be "apprentice scientists" who use the grid as their professional partners do, we start with the emerging grid infrastructure being developed by GriPhyN and iVDgL [7]. We build our learning environment on top of an existing grid architecture used by scientists. We do not attempt to build new grid-based services for the learning environment such as the Learning Flow grid services in the EleGI project.

From a learning perspective, we base our work on twelve years of experience with the LInC Program, which has developed *demonstrators* that integrate best uses of non-grid technology with inquiry-based teaching and learning. The research base comes from the U.S. Department of Education's North Central Regional Educational Laboratory [8].

### 3.1. Cosmic Ray Collaboration

In the Cosmic Ray Collaboration students experience the environment of scientific collaborations in a series of investigations into high-energy cosmic rays. The collaboration is a student-led, teacher-guided project. Schools with cosmic ray detectors can upload data to the server. A virtual data portal (Figure 2) enables students to share these data and associated analysis code with students at other schools whether those schools have their own cosmic ray detectors or not.

Students can pose a number of questions and then design an investigation and analyze the data for answers. Some answers are new to students but well-answered by physicists. These include the muon lifetime, rate of cosmic ray arrival and the source of low-energy air showers. However, the origin of the highest-energy cosmic rays is an open question—several experiments are actually exploring it now. [9, 10] Students may be able to contribute data to these efforts.

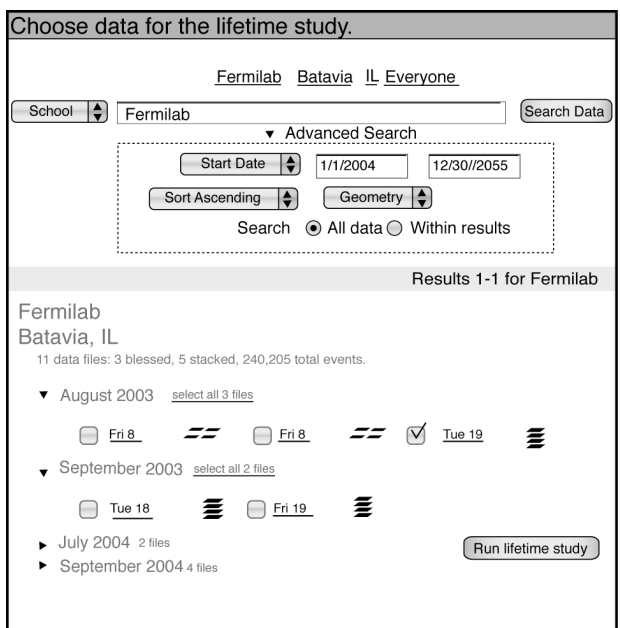

Figure 2. Student interface for searching and selecting uploaded data files for later analysis

Students interpret their data (Figure 3) and post results in online posters or papers. Associated with the posters are comment areas. Students can explore data provenance, comment on the work of others and respond to peer feedback. Because of this feedback they may modify their investigation, test the investigations of others, or design new investigations. In other words, they are *doing science* from a website that uses grid tools and techniques just like their professional colleagues on US ATLAS and USCMS.

### 3.2 Website support - student pages

The collaborative learning environment is supported by a website divided into two major sections: stand-alone student pages and teacher pages. Because students are novice researchers, the website must provide scaffolding to help them navigate a scientific investigation. For example, the first time a student team uses the website after the introductory animation, they are directed through two pages that welcome them to the collaboration, show an example of the end product and provide tips for using the website before going to the study guide, the primary research management tool. The guide presents a list of milestones that, if followed, will result in a complete investigation. The study guide corresponds to what the learning community calls *workflow*. In this paper we use *workflow* in the computer science sense, referring to the flow of data through a set of calculations.

Associated with each milestone is a resource with references and access to a team e-Logbook. A reference page provides links to tutorials for each of the studies, animations, other URLs, access to experts and contact information for all groups in the collaboration. The main navigation bar is consistent throughout the student pages. Most links on the navigation bar also have a submenu for major resources in that section. A pop-up glossary helps students increase their vocabulary. Other tools help with various data analysis and management functions and with poster preparation.

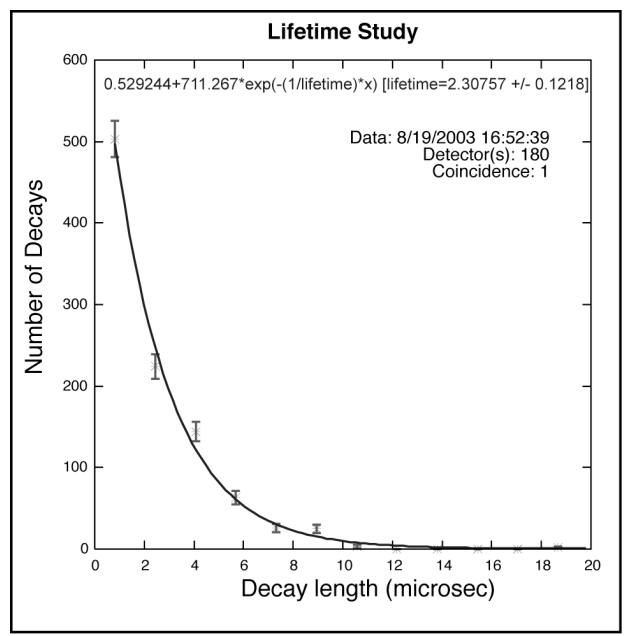

Figure 3. The result of running the lifetime study. Exponential signature provides evidence of muons decaying in scintillator plastic.

The QuarkNet/Grid project is designed for student research teams which is, in itself, a collaborative process. The website promotes asynchronous collaboration among student research groups and between students and experts. Features include a common searchable database, data validation, accessible data files, reproducible computer workflow, comment sections on data files, graphs and posters, and access to experts.

### 3.3 Website support - teacher pages

Collaborative project-based learning requires a skilled teacher/facilitator who understands the importance of each step of the project and allows time for students to move at their own pace through the investigation. With a website for instructional delivery, a teacher needs online tools to manage her classroom. Teacher pages [11] function somewhat like the teacher's edition of a textbook with a general introduction to the project. This includes a list of prior knowledge and skills, learner outcomes, information aligning the student work to U.S. National Science Education Standards, teaching strategies, a site index and more. Currently, the site is designed for QuarkNet teachers who have already participated in staff development programs and need limited assistance with cosmic ray content and student research techniques.

The e-Logbook is a primary tool to track, assess and guide student work because the logbook is linked to the study guide. For example, a teacher can look at and grade all of her research team descriptions of cosmic rays. Or she can look at the notes of a particular team to track progress or see if the students have missed resources that would help them over a trouble spot. A teacher can write comments to the logbooks. In addition, teachers have the option of using online pre- and post-tests of science content and data-handling skills, and a printable rubric to assess achievement of learner outcomes.

## 4. Implementation

The QuarkNet/Grid Collaborative Learning e-Lab is based on the GriPhyN Virtual Data System and the Chiron portal. This section describes the system architecture, the features of the virtual data system and the Chiron portal that are central to our e-Lab and then their specific uses in the QuarkNet/Grid e-Lab.

### 4.1 System architecture

The web-based e-Lab consists of an Apache Tomcat web server with JavaServer Pages (JSP), the GriPhyN Virtual Data System API and libraries (jar files), backend databases for the virtual data system, customized application software, GraphViz for graph visualization and workflow engines for local execution. We expect to add grid execution including Condor-G for grid job submission and execution. The e-Lab requires the Chiron portal as one website and an additional web portal for the e-Lab. Students access the site with browsers that support SSL for security, DHTML and Javascript for interactivity and cookies for session tracking. This provides a platform-independent environment to serve students worldwide.

### 4.2 The Virtual Data System

The GriPhyN Virtual Data System is central to the implementation of the QuarkNet/Grid Collaborative Learning e-Lab. It addresses the need for scientists worldwide with data-intensive applications to share data, data products, and resources needed to create their data products. This system supports the definition of recipes to derive data products with the Virtual Data Language (VDL) and the storage of metadata describing data, data products and recipes in a Virtual Data Catalog (VDC). The metadata and query functionality facilitate the discovery and reuse of resources so scientists can build on each other's work. Requests for data products and recipes can be mapped transparently into computation and/or data access to grid resources worldwide.

### 4.3 Chiron portal

Built on top of this virtual data system is the Chiron portal [12] developed at the University of Chicago by Mike Wilde, Yong Zhao, et al. It provides a web interface to the Java API and services provided by GriPhyN virtual data toolkit. Using a web browser, users can define and execute workflows, publish their data products, share them with other users, and discover and build on the work of other users (Figure 4). Workflows are executed on the local server or in the grid environment.

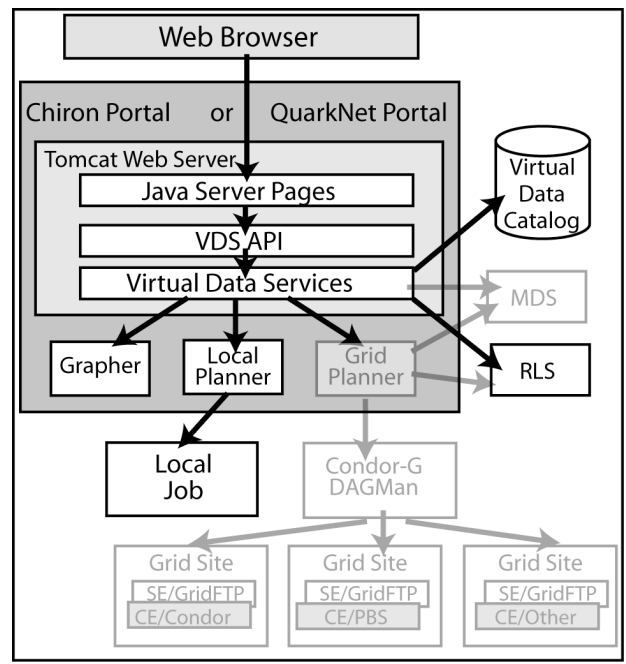

**Figure 4. Chiron/QuarkNet architecture. Light gray items exist in Chiron and are in the QuarkNet development pipeline.**

The Chiron portal itself is general purpose and not tied to any particular scientific application. It provides an environment for a community of users to define the recipes needed for specific scientific applications. The recipes include the definitions of specific data analysis procedures both as transformations (TRs), the counterpart of a function definition, and derivations (DVs), the counterpart of function calls with specific arguments to be used in the execution of the TR. These recipes are represented by text or XML and stored in a database.

**4.3.1 Metadata.** The Chiron portal supports entering metadata, information about the data and definitions, into the VDC both manually and automatically. Metadata is available for datasets, transformations, transformation parameters, derivations, and compound TRs (TRs that reference other TRs). Users can load and search metadata that consists of multiple name, type and value tuples (attribute_name, type, attribute_values) associated with a metadata object (e.g., a logical file name). The values can be integer, float, string, date or boolean. A metadata query language allows users to search for TRs, DVs and logical files with specific attributes. These queries are translated into SQL or XQuery, submitted to the backend database system and returned to the user in an accessible format.

**4.3.2 Provenance.** Provenance provides the audit trail for the derivation of a data product, for example, the transformations and time of their execution, the specific datasets and input parameters (Figure 5). Every time a derivation is executed, a workflow is recorded. Provenance supports the ability of users to re-execute derivations or to modify derivations and provide related data products.

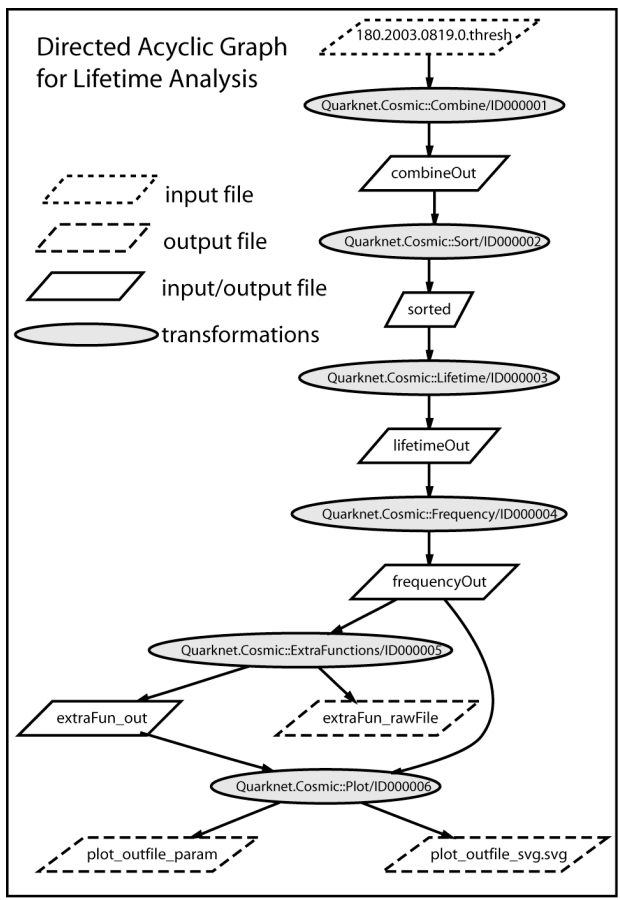

**Figure 5. Workflow for lifetime study**

**4.3.3 Job submission.** Jobs can be submitted to either the local planner to run on the local machine or sent to the grid planner which harnesses grid resources worldwide. The local planner converts the workflow definition into a set of shell scripts to invoke the programs on the local machine. Users use the Chiron portal to upload these applications.

## 4.4 The QuarkNet/Grid portal

Developers of the QuarkNet portal have provided an application-specific web portal that complements and builds on the Chiron portal (Figure 6). It integrates virtual data mechanisms by calling APIs that encapsulate specific Chiron portal functions. It is user-friendly and provides interfaces ("skins") that reflect the needs of the QuarkNet/Grid e-Lab. It hides the details of workflows, transformations and derivations from the high school students and teachers. Developers use the Chiron portal to define transformations and compound transformations in VDL and upload applications (analysis code written in Perl for cosmic ray data). They define specific derivations in the Chiron portal to test that the transformations work in this environment. The Chiron portal displays the transformations and derivations, the data products, DAGs (graphs of the workflow), and all metadata associated with these, including the annotation of the parameters used in the transformations.

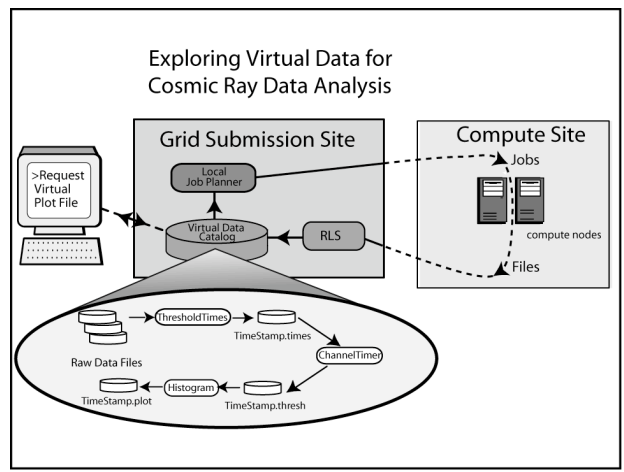

**Figure 6. QuarkNet portal architecture, including data and control paths**

**4.4.1 JavaServer Pages.** JavaServer Pages provide dynamic pages for the students to: upload their data, search for datasets, use forms to enter analysis parameters, analyze their data and publish their work as posters. Each analysis displays a resultant plot generated by Perl code and gnuplot, a graphics package. Students have the option of saving these plots and their accompanying metadata, DAGs and workflows. The JavaServer Pages include calls to middleware to create derivations "on the fly" that are transformed into shell scripts and executed on the local machine to generate the data products. Students use the data products in the online posters they publish.

**4.4.2 Use of metadata.** Metadata is critical to the implementation of the portal. Students can search for, view, and annotate datasets, resultant plots, and posters.

**Figure 7. A poster describing the work done in a lifetime study. The figures cited in the poster are the same as Figures 4 and 6 in this paper.**

All searches involve underlying metadata queries, and students can look at the metadata associated with each. The grid middleware includes APIs to allow setting metadata from JavaServer Pages. Students can annotate data, plots and posters of other students by adding comments to each. Comments are stored as metadata for the data, plot or poster. Comments as metadata provide a way for students to communicate and collaborate.

Students can look at figures associated with posters and find the DAGs associated with data plots. They will soon be able to recreate any plot made by other student research groups and then manipulate input parameters to make related plots. This functionality for discovery allows them to build on each others' work as scientists do.

**Table 1. The metadata for the object represented by Figure 8**

| META TAG | VALUE |
| --- | --- |
| author: | Thomas Jordan |
| | Liz Quigg |
| | Eric Gilbert |
| | Bob Peterson |
| city: | Batavia |
| date: | 2004-11-1000:00:00.0 |
| group: | Fermilab |
| name: | poster_decays.data |
| plotURL: | Users/…/fermigroup/cosmic/plots |
| project: | Cosmic |
| school: | Fermilab |
| state: | IL |
| teacher: | Jordan |
| title: | Possible Particle Decays |
| type: | Poster |
| year: | AY2004 |

The Cosmic Ray Collaboration website requires lots of "help" links and background. Metadata provides an excellent way to integrate this content into the site. To understand the analysis inputs students see on their web pages, students access the annotations of input parameters entered by the developer using the Chiron portal. The metadata implementation is hidden from the students; to them it is just another link to a pop-up web page. Developers have defined two classes of documentation, glossary items and references. References are short web pages that provide background information for students in their investigation. They can include links to other web pages within or outside the portal. All glossary items and references are associated with logical file names in the VDS starting with Glossary_ and Reference_, respectively. Both have metadata annotations including the name, type (glossary or reference) and the description of the item. Metadata can include HTML, so the definitions of glossary items are stored as text metadata that can be displayed to students in web pages. Web pages available only to content providers for the Cosmic Ray Collaboration allow defining glossary items and references. Summary pages of all the glossary items and references are available for the students.

**4.4.3 User database.** We have developed a user database in PostgreSQL to store information about the research groups, their teachers, their locations, their roles, questions for pretests, student responses to the pre- and post-tests and the content of the e-Logbook for each research group and teacher. Web pages provide registration of stu-dent research groups and the administration of the pre- and post-tests used for program evaluation and student assessment.

## 4.5 Lessons learned from the QuarkNet grid implementation

The implementation of the QuarkNet/Grid portal has provided valuable insight into what other e-Labs will require, which components are e-Lab-independent, and which are e-Lab-specific. Independent components include basic website structure, registration, login, data interface, some transformation, e.g., the histogram transformation, application of transformations to data, documentation of variables used by science code, implementation of glossary and references, authoring posters, comments on posters and data

It has also provided information on what additional functionality is needed in the Chiron portal to make it more useful to students and scientists. For example, currently the Chiron portal, as opposed to the QuarkNet/Grid portal, does not display or allow searches of datasets. The interface to define transformations is difficult to use and could be improved with a graphical interface. It is not as user-friendly as average high-schoolers require for their work. However, the Chiron portal and its accompanying APIs empower the developers. It provides an authoring system and interface to the Virtual Data System required to build the QuarkNet/Grid e-Lab website.

Professional development is an important component of our program. Our experience is currently limited to teachers that understand the content embedded in the QuarkNet portal. We have three modes of instruction for these teachers. We focus on the web portal in all three modes. In large workshops (one-week, 25 participants), we model the practices and habits of scientists. Teachers join their peers in research groups, meet the problem at hand and work out the solution using the portal. In addition to modeling behaviors of scientists, the teachers use the portal as their students do. In smaller, one-day workshops we provide more guidance to the teacher's exploration of the portal. These shorter sessions leave less room for exploration, and we highlight important features of the portal. We also visit teachers and students for one-to-few coaching in the use of the portal. We learn as much about the interface as users do about the science in these sessions.

It is too early to determine if the QuarkNet/Grid setting enhances learning or the role that technology plays in enhancing learning. However, we gain valuable information from formative evaluation as we iterate the portal. Teacher focus groups use the portal as if they were students to identify showstoppers or sticking points. They identify areas for additional classroom management tools and point out information needed on the teacher pages.

Our outside evaluator provides input based on her experience using the portal with students. This "product testing" continues to be an important development step.

## 5. Next steps

The QuarkNet portal provides a solid foundation upon which students participate in real science collaboration. We have thus far focused primarily on delivering a highly usable and appealing science collaboration for a specific audience of students. Looking ahead, we envision and have begun developing numerous enhancements that will deliver richer science collaborations to students.

### 5.1 Student access to transformations

Advanced students (primarily in computer science classes) will have access to the Chiron portal if they would like to understand the underlying functionality, to combine the available transforms in new ways and to write their own transforms. For example, students with appropriate permissions could write and upload new Perl code through the Chiron portal. By combining the new code with existing atomic transformations, the students make new analysis tools. These new analysis tools could then become public to the collaboration.

### 5.2 Data simulations

Web experiments exist that enable students to interactively investigate the concept of background in data. For example, Ultra Electronics interactive demos [13] give students the ability to hear the noise of a jet before and after noise reduction. Students must make careful determinations of similar background in cosmic ray data. They can use data from another detector to see how often their signal is mimicked. They must then subtract this background from their measurements. We will develop tools to facilitate this measurement and subtraction.

### 5.3 Additional e-Lab topics

Currently, we are refactoring our original code design as a first step to creating a developer toolkit for other science experiments, like CMS. We will convert our JSP-only design into an Object-Oriented JAR file hierarchy. In this way we can provide documentation and an organized approach to creating e-Labs. The VDS code is open source, and the e-Lab toolkit will be open source.

### 5.4 Farming out calculations and data to the grid

We have reached a critical point in the QuarkNet/Grid project where a single computer is no longer enough to analyze the datasets students have accumulated. Currently, a student makes a request to analyze a set of files and waits synchronously for the result to return. In the beginning of our project, this paradigm made sense.

However, if a student chooses to look for a shower at three schools over a period of three days, she can expect to wait about an hour for the computation to complete; or longer for very large datasets. We plan to build an asynchronous user interface for web-based grid job submission. As a result of this requirement, QuarkNet is pushing VDS feature development. We contribute our solutions back into the main VDS development line.

### 5.5 Support for teachers without detectors or outside QuarkNet

While the current project focuses on students in classrooms with QuarkNet teachers, we envision a much larger collaboration including students whose teachers participate in other collaborations such as the Cosmic Ray Observatory Project (CROP) [14] and Washington Large Area Time Coincidence Array (WALTA) [15], and teachers not affiliated with such projects. Even within QuarkNet, some teachers will not have access to a classroom detector. This will require us to develop more background material for students and materials as well as professional development activities for teachers. As a first attempt towards including a greater number of teachers, we will include QuarkNet teachers without detectors. Students from these classrooms will participate by examining, analyzing and commenting on other students' data. As a long-term goal, we strive to include an increasing number of online resources that enable teachers to pursue the project without training from the national QuarkNet staff. This will increase the size of the collaboration. For example, we have doubled the number of resources we offer since December 2004.

### 5.6 Evaluation

An outside evaluator is designing an assessment to determine the effectiveness of using grid techniques and tools to support learning. The evaluation is divided into two phases. Currently, we are completing Phase I—testing and improving the student web interface and grid tools. Activities involve iterative testing using a protocol to watch teachers and students use the site and improving the site. They also include an assessment of prior knowledge and skills needed to use the site. After the site is stable, we move to Phase II—an assessment of learning physics content and using scientific data. Activities include building online pre- and post-tests, student observations, "think aloud" interviews, review of student posters and e-Logbooks, and a discourse analysis on student use of communication and collaboration tools.

### 6. Acknowledgments

The authors would like to thank Nick Dettman, Evgeni Peryshkin, Robert Peterson and Yun Wu. These members of the development team made substantial contributions to the portal. This work is partially funded by the National Science Foundation and the U.S. Department of Energy Office of Science.